\documentclass[aps, prl, reprint, superscriptaddress, twocolumn, showkeys, amsmath, amssymb, longbibliography, floatfix]{revtex4-1}

\usepackage[dvipsnames]{xcolor}

\usepackage[english]{babel}
\usepackage{
amsmath, 
amssymb, bbm, mathrsfs, bm, braket, color, graphicx, comment, amsfonts, dsfont}
\usepackage[colorlinks, linkcolor=blue, citecolor=blue, urlcolor=blue]{hyperref}
\usepackage{hyperref}
\usepackage[mathscr]{euscript}
\usepackage{bm}

\newcommand{\pd}{{\phantom\dag}}

\begin{document}

\title{Tuning the topological winding number by rolling up graphene}

\author{Ying-Je Lee}
\affiliation{Department of Physics, National Cheng Kung University, Tainan 70101, Taiwan}
\affiliation{Center for Quantum Frontiers of Research and Technology (QFort), National Cheng Kung University, Tainan 70101, Taiwan}

\author{Yu-An Cheng}
\affiliation{Department of Physics, National Cheng Kung University, Tainan 70101, Taiwan}
\affiliation{Center for Quantum Frontiers of Research and Technology (QFort), National Cheng Kung University, Tainan 70101, Taiwan}

\author{Yu-Jie Zhong}
\affiliation{Department of Physics, National Cheng Kung University, Tainan 70101, Taiwan}
\affiliation{Center for Quantum Frontiers of Research and Technology (QFort), National Cheng Kung University, Tainan 70101, Taiwan}

\author{Ion Cosma Fulga}
\email{i.c.fulga@ifw-dresden.de}
\affiliation{Institute for Theoretical Solid State Physics, IFW Dresden, Helmholtzstraße 20, 01069 Dresden, Germany}
\affiliation{W\"{u}rzburg-Dresden Cluster of Excellence ct.qmat, Germany}

\author{Ching-Hao Chang}
\email{cutygo@phys.ncku.edu.tw}
\affiliation{Department of Physics, National Cheng Kung University, Tainan 70101, Taiwan}
\affiliation{Center for Quantum Frontiers of Research and Technology (QFort), National Cheng Kung University, Tainan 70101, Taiwan}
\affiliation{Program on Key Materials, Academy of Innovative Semiconductor and Sustainable Manufacturing, National Cheng Kung University, Tainan 70101, Taiwan}

\date{\today} 

\begin{abstract}
Nanoscrolls, radial superlattices formed by rolling up a nanomembrane, exhibit distinct electronic and magneto-transport properties compared to their flat counterparts. 
In this study, we theoretically demonstrate that the conductance can be precisely enhanced $N$ times by rolling up graphene into an $N$-turn nanoscroll and applying a longitudinal magnetic field. 
This tunable positive magnetoconductance stems from the \emph{topological} winding number which is activated in a carbon nanoscroll with magnetic flux and its maximum value purely increases with the \emph{scroll} winding number (the number of turns). 
By integrating material geometry and topology, our work opens the door to artificially creating, customizing, and designing topological materials in rolled-up graphene-like systems.
\end{abstract}

\maketitle

\textcolor{blue}{\textit{Introduction}} --- 
Tuning material topology through nanoarchitecture is gathering great attention for its profound implications in fundamental science and its potential to advance nanotechnology applications. 
Recently, the Chern number in two-dimensional (2D) periodic systems has been artificially increased to five by constructing sophisticated lateral heterostructures, including magnetic topological insulators~\cite{Nature2020} and multilayered graphene systems~\cite{Science2024}. 
This development opens new avenues for enhancing the quantum anomalous Hall effect artificially~\cite{Nature2020, Science2024}. 
In this context, mastering the control of topological winding numbers in one dimension will be crucial for further enriching the exploration of material topology across various dimensions.

In addition to designing lateral heterostructures, rolling a nanomembrane into a nanoscroll introduces unique electronic properties that offer diverse functionalities and a high application potential. 
Nanoscrolls exhibit distinct transport behaviors compared to their flat counterparts due to the spiral cross-section, inter-winding coupling, and open boundaries. 
For example, van der Waals nanoscrolls are characterized by impressive electronic transport properties~\cite{nanoscroll2009, cui2018, zhao2021, zhong2022,zhu2024}. 
Both carbon nanoscrolls and phosphorus nanoscrolls are noted for either high conductivity~\cite{Sch2011} or high carrier mobility~\cite{wang2020, wang2018}. 
The nontrivial geometric structure of nanoscrolls~\cite{wang_2024}, functioning as a radial superlattice, may give rise to rich transport properties, as highlighted in pioneering studies.

\begin{figure}[h!]
\centering
\includegraphics[width=1\linewidth]{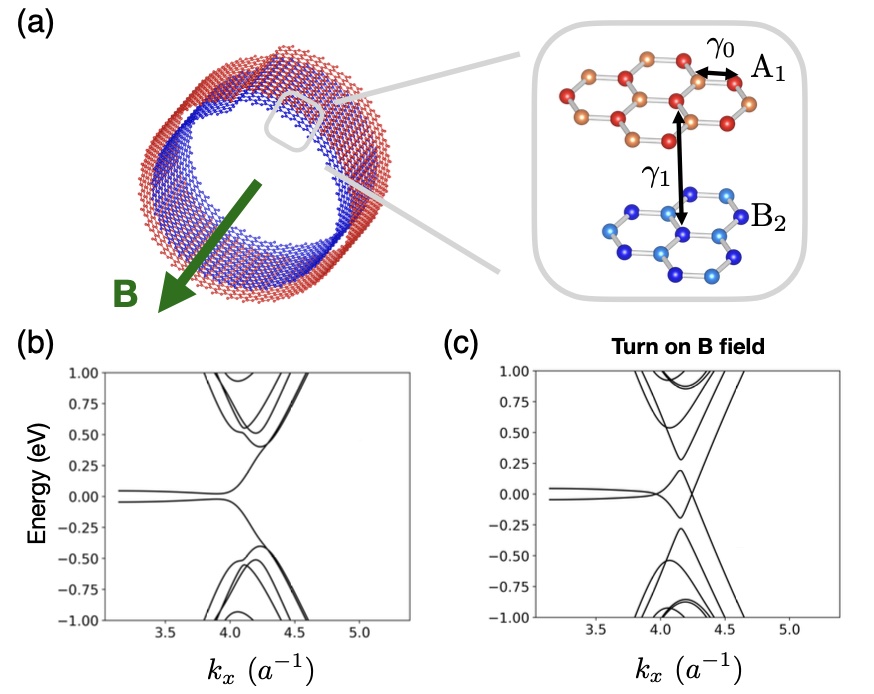}
\caption{
Band structures near the high-symmetry K point for a two-turn carbon nanoscroll with an arc-length of 10.2 nm [panel (a)], computed (b) without and (c) with an axial magnetic flux of magnitude $\phi= 0.5~\phi_0$, where $\phi_0$ is the magnetic flux quantum.
Introducing a magnetic field in the two-turn nanoscroll leads to two band crossings at $E=0$.
The inset in panel (a) shows the tight-binding model used in our simulations.
}
\label{fig1}
\end{figure}

In this work, we study the quantum states in carbon nanoscrolls and find that their bands will cross at the Fermi level ($E=0$) when a magnetic field with proper strength $B$ is applied along the nanoscroll core axis (see Fig.~\ref{fig1}). 
Moreover, the number of band crossings is directly determined by the number of nanoscroll turns, resulting in an $n$-turn nanoscroll with a magnetoconductance near $E=0$ of:
\begin{equation}\label{eq:GofB}
\triangle G = G(B)-G(0) =N \times 8\frac{e^2}{h},
\end{equation}
where $e$ represents the electron charge, and $h$ is the Planck constant. 
The turn number $n$ is not required to be an integer and $N=\lfloor n \rfloor$ is a floor mapping $n$ to a largest integer  satisfying $N \leq n$. 
Moreover, the coefficient $8=2^3$ in Eq.~\eqref{eq:GofB} is contributed by the state degeneracy of two spins, two valleys and the band crossing. 
We confirm that a topological winding number, increasing precisely with the number of turns in the nanoscroll with magnetic flux, drives this positive magnetoconductance.

\textcolor{blue}{\textit{Model}} --- 
To investigate the energy bands of carbon nanoscrolls with inter-winding electronic couplings and applied magnetic flux, we start by rolling a graphene sheet to maintain the zigzag edge boundary of the carbon nanoscroll, ensuring that its boundary is a cross-section arranged in a rhombohedral stacking \cite{Warner2012Jun} configuration (ABC-stacking) of carbon atoms.   
We begin with the spinless tight-binding model of a zigzag-type graphene nanoribbon with nearest-neighbor hopping 
 \begin{align}
H_{\rm rib}=\gamma_0\sum_{\langle {\bf r, r'}\rangle} a_{\bf r\phantom{}}^{\dag}b^{\pd}_{\bf r'}+{\rm h.c.},
 \label{graphene-Hamiltonian}
 \end{align}
where  $\gamma_0$ is the value of intralayer coupling and $a$ ($b$) denote the A-site (B-site) annihilation operators, with A and B labeling the sublattices. 
Here, ${\bf r} = i, j$ with $ i $ referring to the $i^\text{th}$ unit cell along the axial direction, and $ j$ representing the $j^\text{th}$ AB-pair within the ribbon's open boundaries. A nanoribbon with a width of $m$ unit cells is simulated by imposing the condition $j \leq m $.

When the nanoribbon rolls up into an $n$-turn carbon nanoscroll,
the full Hamiltonian $H$ can be built up as $H=H_{\rm rib}+H_{\rm inter}$ where
\begin{equation}
H_{\text{inter}} = \gamma_1 \sum_{j=1}^{j_{\rm max}} a_{i,j}^\dagger b^{\pd}_{i,j+\lceil\frac{m}{n}\rceil-1} + \rm{h.c.}
 \label{scroll-interlayer-coupling}
\end{equation}
accounts for the interlayer A-B site hopping within the nanoscroll. The interlayer hopping strength is denoted by $\gamma_1$, and the net number of interlayer hopping is given by $j_{\rm max}=m+1-\lceil m/n \rceil$. Here the ceiling function maps $m/n$ to the smallest integer greater than or equal to $m/n$. For example, a two-turn nanoscroll with 100 unit cells along its arc can be modeled by setting $n = 2$, $m = 100$, and $j_{\rm max}=51$  in Eqs. (\ref{graphene-Hamiltonian}) and (\ref{scroll-interlayer-coupling}).

Our numerical simulations of carbon nanoscrolls are based on the kwant package~\cite{Groth2014Jun}, setting an intralayer coupling $\gamma_0 = 3.16$ eV, an interlayer coupling $\gamma_1= -0.381$ eV~\cite{Kuzmenko2009Oct, McCann2013}, and the lattice constant $a=2.46 $ \AA~\cite{Saito1998Jul}.
Additionally, we include the effect of a magnetic flux applied along the nanoscroll axis. 
We model it by introducing an effective intralayer coupling parameter with the Peierls phase~\cite{Peierls1933Nov, Do2022Jun, Lesser2020Jun} (see Supplemental Material (SM) \cite{supp}).

In the following, we investigate the topological magnetotransport properties of the carbon nanoscroll by simulating a system with an arc length of 10.2 nm. 
We have verified that our results remain both qualitatively and quantitatively consistent for nanoscrolls with longer arc lengths, up to 51 nm (see SM \cite{supp}).

\textcolor{blue}{\textit{Band structure}} --- 
For an infinitely long two-turn carbon nanoscroll with arclength 10.2 nm, the band structure without and with the applied axial magnetic field is shown in Figs.~\ref{fig1}(a) and \ref{fig1}(b), respectively.
Focusing on the energy bands near the projection of the high-symmetry K point, which is also the Dirac point in monolayer graphene, we can find that a band gap of about  $0.04$ eV opens up in the nanoscroll, as shown in Fig.~\ref{fig1}(a). 
And yet, as shown in Fig.~\ref{fig1}(b), the band gap in the low-energy region vanishes and two energy bands cross at $E=0$ when a magnetic flux equal to half of the magnetic flux quantum is applied.

\begin{figure}[t]
\centering
\includegraphics[width=1\linewidth]{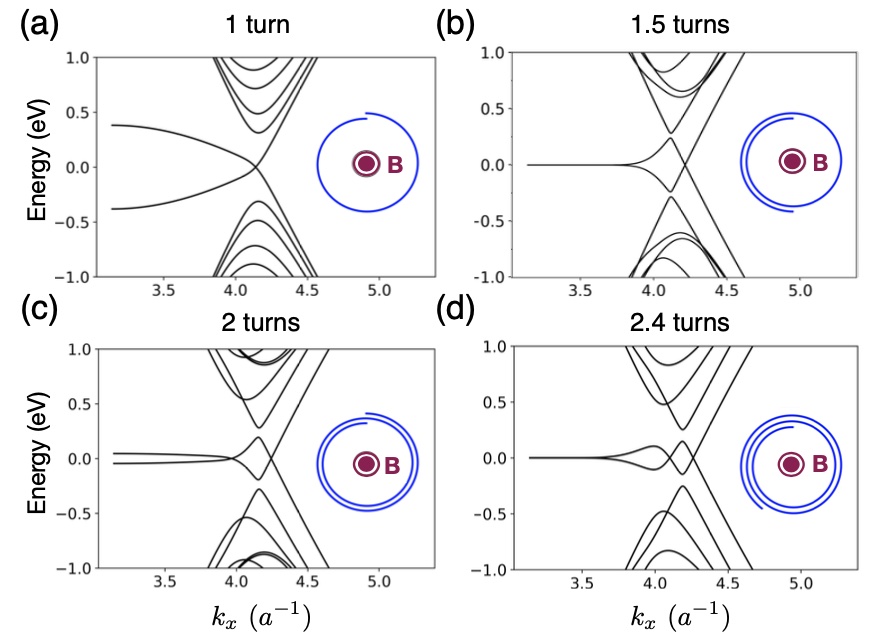} 
\caption{
The band structures near the projection of the high-symmetry K point of carbon nanoscrolls are examined as a function of longitudinal momentum with a fixed arc-length of 10.2 nm, varying the turn number from 1.0 to 2.4, while maintaining the external magnetic flux magnitude at $ \phi = 0.5~\phi_0$. 
The energy bands of carbon nanoscroll are shown for (a) 1.0 turn, (b) 1.5 turns, (c) 2.0 turns, and (d) 2.4 turns.
}
\label{fig2}
\end{figure}

By comparing Figs.~\ref{fig1}(a) and \ref{fig1}(b), we observe that the additional energy band crossings at $E=0$ are driven by the applied magnetic flux. 
In Fig.~\ref{fig2}, 
we roll up a graphene ribbon into a carbon nanoscroll, maintaining its total width, while varying the number of turns from approximately 1.0 to 2.4.
Additionally, we set the magnetic flux to a half magnetic flux quantum $\phi/\phi_{0}=0.5$ to explore the relationship between the number of band crossings and the number of turns of carbon nanoscrolls. 
As shown in Figs.~\ref{fig2}(a) to \ref{fig2}(d), our findings indicate that the number of band crossings at low energy $E=0$ increases with the number of turns (scroll windings) of carbon nanoscrolls. 
Figures \ref{fig2}(b) and \ref{fig2}(d) also reveal that for non-integer turn configurations, the band structure exhibits a flat band at $E=0$, akin to that of a zigzag graphene ribbon with edge states at $E=0$. 
However, this feature disappears for configurations with an integer number of turns, as seen in Figs.~\ref{fig2}(a) and \ref{fig2}(c). 
We will show that the number of scroll windings is an adjustable parameter that controls the number of energy band crossings and the occurrence of the flat ribbon states.

\textcolor{blue}{\textit{Topological winding number} }--- 
To understand why the number of scroll windings increases together with the number of band crossings at $E=0$, we investigate the topological winding number of electronic states in the carbon nanoscroll.
We begin by converting the Hamiltonian $H$ of the infinitely long nanoscroll from its operator form to matrix form
\begin{equation}
H^{\prime} = 
 \begin{pmatrix}
  0 & h(k_x, \phi)  \\
  h^{\dagger}(k_x, \phi) & 0    
 \end{pmatrix},   \label{chiral-Hamiltonian}
\end{equation} 
based on the set separating A and B sites: $\lbrace A_1,A_2,...,A_{\rm max}, B_1, B_2,...,B_{\rm max} \rbrace$.
Notice that the Hamiltonian matrix is block off-diagonal, as a consequence of sublattice symmetry.
For a two-turn nanoscroll with 16 lattice sites in the cross section, for instance, its off-diagonal part $h(k_x, \phi)$ is:
\begin{equation}
h(k_x, \phi) = 
 \begin{pmatrix}
  t_1 & 0 & 0 &\gamma_1 & 0 & 0 & 0 & 0   \\  
  t_2p^{\ast} & t_1^{\ast}  & 0 & 0 & \gamma_1 & 0 & 0 & 0  \\  
  0 & t_2  & t_1 & 0 & 0 & \gamma_1 & 0 & 0   \\  
  0 & 0  & t_2 & t_1^{\ast} & 0 & 0 & \gamma_1 & 0    \\  
  0 & 0  & 0  & t_2 & t_1 & 0 & 0 & \gamma_1 \\   
  0 & 0 & 0 & 0 & t_2p^{\ast} & t_1^{\ast} & 0 & 0   \\  
  0 & 0 & 0 & 0 & 0 & t_2 & t_1 & 0  \\ 
  0 & 0 & 0 & 0 & 0 & 0 & t_2 & t_1^{\ast}   
 \end{pmatrix},    \label{scroll-Hamiltonian-diag}
\end{equation}
where parameters in $h(k_x, \phi) $ are $t_1=\gamma_0(e^{ik_{x}a}+1)$, $t_2=\gamma_0$, $t_2p=\gamma_0 e^{i2\pi \phi/\phi_0}$, $t_2p^{\ast}=\gamma_0 e^{-i2\pi \phi/\phi_0}$, and
$t_1^{\ast}=\gamma_0 (e^{-ik_{x}a}+1)$.  The value $\phi_0$ is as the magnetic flux quantum and the detailed derivation is provided in the SM \cite{supp}.

The block off-diagonal basis enabled by sublattice symmetry enables us to use $h(k_x, \phi)$ in order to determine the topological winding number associated to the Hamiltonian $H^{\prime}$ (see SM \cite{supp}), which reads~\cite{Zak1989, Ryu2002, Chiu2016Aug, Asboth} 
\begin{equation}
\nu=\dfrac{1}{2\pi i} \oint \dfrac{d}{d \phi} \ln  \det\left[ h\left(k_x, \phi\right)\right]
\label{winding_number}
\end{equation}
Note that Eq.~\eqref{winding_number} is a nonstandard form for the winding number, which usually takes the form of a momentum integral.
Here, in contrast, the momentum is kept fixed and the integral is performed along a different periodic variable, the flux threading the nanoscroll.
In essence, the momentum $k_x$ and the flux $\phi$ can be seen as two periodic variables defining a ``synthetic'' two-dimensional (2D) Brillouin zone.

\begin{figure}[t]
\centering
\includegraphics[width=1\linewidth]{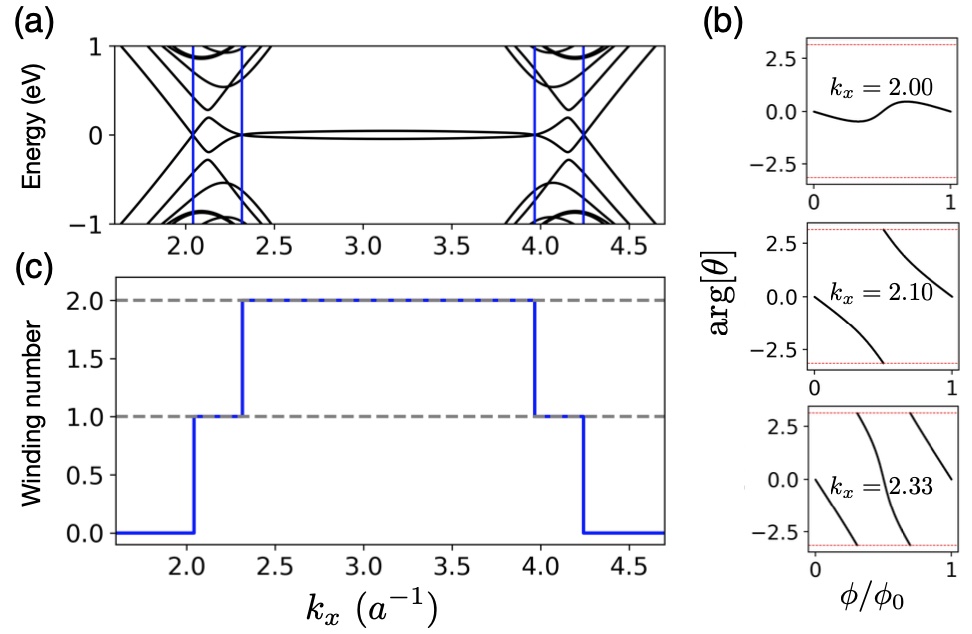} 
\caption{
(a) The energy band of the entire Brillouin zone for a two-turn carbon nanoscroll with an arc-length of 10.2 nm. 
A magnetic flux with magnitude $\phi = 0.5~\phi_0$ is applied. 
The band crossings are shown with colored vertical lines that show their corresponding momentum positions. 
(b) In regions with different winding numbers, select three momentum coordinate points, $k_x = 2.0$, $2.1$, and $2.33$, and plot the angle diagram of $\det[h(\phi)]$ in the complex plane relative to the origin, $\arg[\theta]$.
(c) Phase diagram of the winding number of $\det[h(\phi)]$ at different momenta. 
}
\label{fig3}
\end{figure}

Figure \ref{fig3}(a) displays the two-valley band structure of the two-turn nanoscrolls under the application of a half magnetic flux quantum. 
We have marked the positions of the momentum $k_x$ corresponding to the energy band crossings in this figure. 
The momentum at these intersections can serve as a critical parameter for distinguishing different topological phases.
To investigate the behavior of the winding number, we track the phase of the natural logarithm of $\det h(k_x, \phi)$ at fixed values of $k_x$.
As shown in Fig.~\ref{fig3}(b), the winding number changes as $k_x$ is swept across the band crossing points, leading to the appearance of different topological phases. 
Different momentum intervals across the entire Brillouin zone of the nanoscroll correspond to distinct topological winding numbers, taking values 0, 1, and 2, respectively [see Fig.~\ref{fig3}(c)].
We conclude that the band crossings are topologically protected.
In addition, we have verified that the band structures calculated for various turns, e.g. as in Fig.~\ref{fig2}, clearly demonstrate that the winding number also increases as the number of turns increases. 
The emergence of more band crossing points signifies the availability of a greater number of transport modes at low energies.

\begin{figure}[tb]
\centering
\includegraphics[width=1\linewidth]{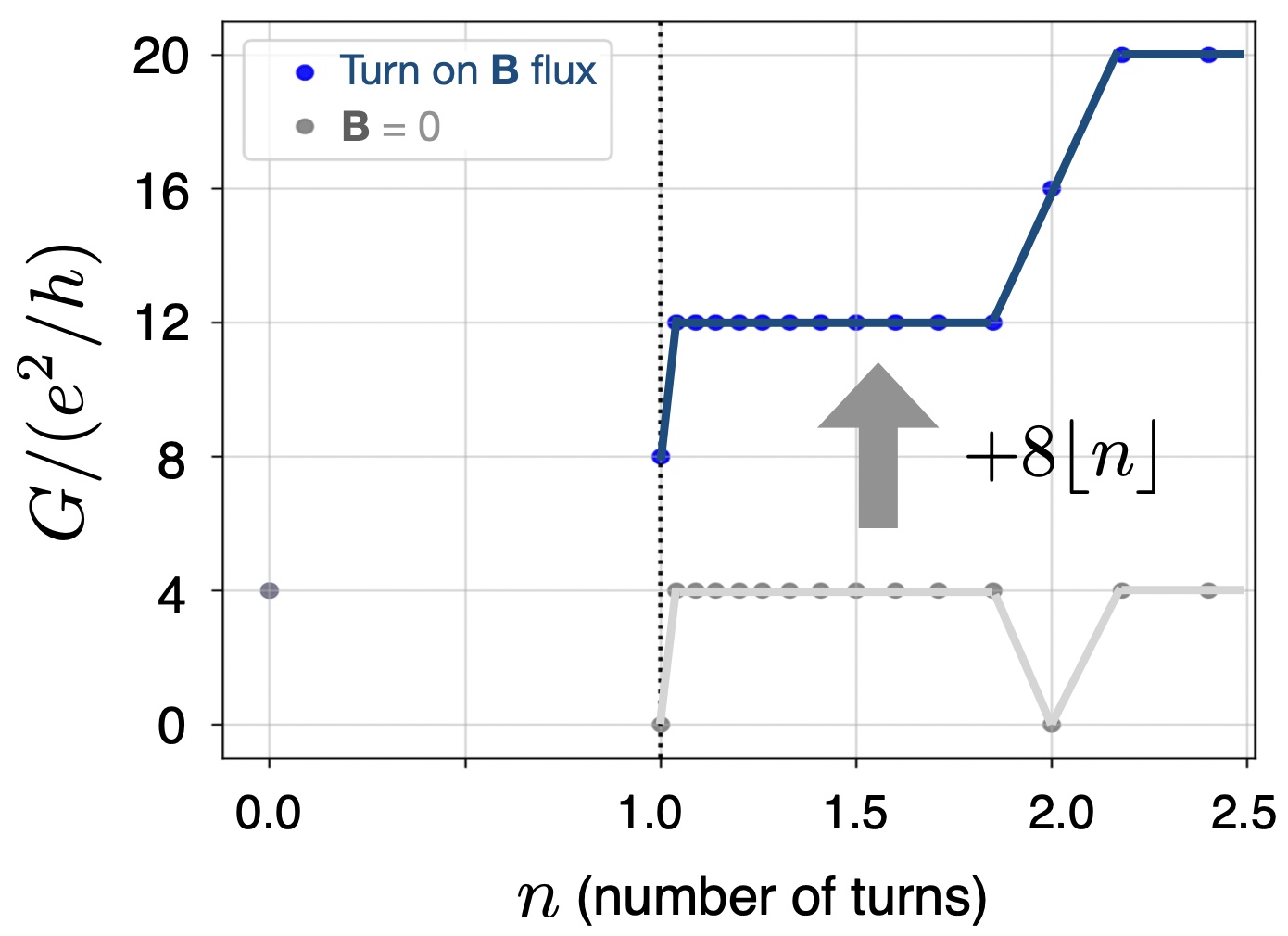}
\caption{
By varying the number of turns from 0 to 2.4, a conductance diagram for carbon nanoscrolls is obtained. 
The results for the nanoscroll with the applied magnetic flux magnitude is $\phi = 0.5~\phi_0$ are denoted in black line which links our solved data. 
The gray line indicates the conductance of the same nanoscroll but in absence of magnetic flux.}
\label{fig4}
\end{figure}

\textcolor{blue}{\textit{Magnetoconductance}} ---
To investigate the impact of band crossings that are tunable in the nanoscroll with magnetic flux, we determine the conductance of a nanoscroll with fixed arclength while increasing its number of turns, as shown in Fig.~\ref{fig4}.
The conductance is obtained at low energy near $E=0$, corresponding to the system with low carrier density.   
We begin with a nanoribbon with zigzag boundaries (zero turns), where the conductance provided by edge states is 4 $e^2/h$, where $e$ is the electron charge and $h$ is the Planck constant.
Here, again, the factor of 4 is due to the valley as well as the spin (which is not explicitly included in our tight-binding model).
For the nanoscroll without a magnetic flux, which retains the same lattice but introduces interlayer coupling due to its rolled-up structure, the conductance switches between 0 and 4 in units of $e^2/h$ as the number of turns transits between the integer and non-integer, as shown by the gray line in Fig.~\ref{fig4}.
These results reflect the coexistence of two edge states in the nanoscroll with open boundaries. The energies of these states split apart from zero when the nanoscroll has an integer number of turns, due to hybridization arising from the direct overlap of the boundaries.

For the nanoscroll with applied magnetic flux $\phi=0.5~\phi_0$, however, the nature of conductance changes crucially -- its value simply increases as a step-like function with the number of turns, as shown by the blue line in Fig.~\ref{fig4}.
The conductance difference between the nanoscroll with and without magnetic flux is as described in Eq.~\eqref{eq:GofB}.
Our results in Fig.~\ref{fig4} demonstrate that the applied magnetic flux can turn on and multiply the conductance in a low-carrier-density nanoscroll with integer and non-integer turns, respectively.

\textcolor{blue}{\textit{Conclusion} }--- 
In sum, we theoretically demonstrate, comparing to the flat counterpart, that the longitudinal magnetoconductance switches on and is noticeably enhanced by rolling up graphene into a carbon nanoscroll.
We establish that this unique magneto-transport stems from the appearance of a topological winding number, the topological index in one dimension, in the nanoscroll with applied magnetic flux.
Furthermore, we find that this enhancement of magnetoconductance, driven by the topological nature, is directly proportional to the number of turns of a carbon nanoscroll.
Our findings indicate that not only the material itself but also the scroll winding are critical and tunable parameters to designing quantum transport in graphene-like nanoarchitectures.
We note that carbon nanoscrolls have been manufactured and investigated for more than a decade, and that they can naturally appear at the edge of graphene monolayers \cite{nanoscroll2009, wang_2024}.
Our results open the way towards connecting the material geometry and topology to create and artificially control the high-disorder resistant and high magneto-transport performance in curved nanoarchectures. 
We hope that the insights and effects of the magnetotranport we have revealed in our work will be observed in experiments.
\section*{Acknowledgements}
We acknowledge the financial support by the National Science and Technology Council (Grant numbers 112-2112-M-006-026-, 112-2112-M-004-007 and 112-2112-M-006-015-MY2) and National Center for High-performance Computing for providing computational and storage resources. C.-H. C. thanks A. Huang for helpful discussions and assistance. C.H.C. thanks support from the Yushan Young Scholar Program under the Ministry of Education in Taiwan. This work was supported in part by the Higher Education Sprout Project, Ministry of Education to the Headquarters of University Advancement at the National Cheng Kung University (NCKU).
I.C.F was supported by the Deutsche Forschungsgemeinschaft (DFG, German Research Foundation) under Germany's Excellence Strategy through the W\"{u}rzburg-Dresden Cluster of Excellence on Complexity and Topology in Quantum Matter--\emph{ct.qmat} (EXC 2147, project-id 390858490).

%

\end{document}